# Recognizing Cancer via Somatic and Organic Evolution

Xiang-Ping Jia[*] and Hong Sun[§]

[*]*College of Science, Liaoning University of Technology, Jinzhou, P. R. China 121001*

[§]*College of Foreign Languages, Liaoning University of Technology, Jinzhou, P. R. China 121001*

[*]*E-mail:* [*jiaxiangping@tsinghua.org.cn*](mailto:jiaxiangping@tsinghua.org.cn)

**Running title:** Understanding the Nature of Cancer

**Abstract**

The fitness of somatic cells of metazoan, the ability of proliferation and survival, depends on microenvironment. In somatic evolution, a mutated cell in a tissue clonally expands abnormally because of its high fitness as normal cells in a corresponding microenvironment. In this study, we propose the *cancer cell hypothesis* that cancer cells are the mutated cells with two characteristics: clonal expansion and damaging the microenvironment through the behaviours such as producing more poison in metabolism than normal cells. This model provides an explanation for the nature of invasion and metastasis, which are still controversial. In addition, we theoretically reasoned out that normal cells have almost the highest fitness in healthy microenvironments as a result of long-term organic evolution. This inspires a new kind of therapy of cancer, which improving microenvironment to make cancer cells lower in fitness than normal cells and then halt the growth of tumours. This general therapy relies on a mechanism differing from chemotherapy and targeted therapy.

**Keywords**: multi-step tumourigenesis, evolution, fitness, metastasis, microenvironment, cancer therapy

## Multi-step process of tumourigenesis

Cancer is an abnormal proliferation of somatic cells caused by accumulating mutations in genes which control cell growth and differentiation (Bertram, 2000; Olopade & Pichert, 2001; Vickers, 2007). It is widely accepted today that tumourigenesis is a multi-step process involving about 4 ~ 6 alterations in DNA of somatic cells (Armitage & Doll, 1954; Cook *et al*, 1969; Nordling, 1953; Weinberg, 2007). The multi-step model is effective in explaining the age-related incidence of cancer, the late onset of most cancer (Knudson, 2001; Weinberg, 2007) and that the formation of tumours requires an extended period of repeated exposure to carcinogens (Peto, 2001). A typical example according with multi-step model is colon carcinoma progresses in multi steps as (i) loss APC leads to hyperplastic epithelium, (ii) DNA hypomethylation leads to early adenomas, (iii) activation of K-*ras* leads to intermediate adenomas, (iv) loss of 18q TSG leads to late adenomas, loss of P53 leads to carcinoma and (v) invasion & metastasis (Vogelstein *et al*, 1988; Vogelstein *et al*, 1989; Weinberg, 2007). In addition, the multi-step model is verified in breast cancer, and chronic myelogenous and acute lymphoblastic leukaemia (Karakosta *et al*, 2005).

The multi-step tumourigenesis follows Darwinian evolution, in which individual cells compete with one another rather than individual organisms compete with one another (Weinberg, 2007). Each stage in this process involves two events (Fig. 1). The first event is that one cell amid a large cell population sustains a specific mutation. The second event is that the mutated cell clonally expands a large number. A special mutation confers on the mutated cell a proliferation and survival advantage compared with those cells lacking this mutation (Weinberg, 2007). Mutagens as ionized radiation cause directly the mutation of cells. Differing from mutagens, must carcinogens as smoking, aflatoxin and asbestos may promote the abnormal proliferation of mutated cell by influencing microenvironment (Fig. 1). Mutation and proliferation are two indispensible and different phases of tumourigenesis in which a normal cell gradually transforms into a cancer cell.

Stephen Paget proposed "seed and soil" hypothesis based on the specificity of target tissue for tumour metastasis more than a century ago (Ribatti *et al*, 2006). In this hypothesis, the formation of metastasis depends on the properties of the tumour cells as well as the permissive role of the environment (Lorusso & Rüegg, 2008). Tumour-associated microenvironment consists of cells, soluble factors, signalling molecules, extra cellular matrix and mechanical cues (Bissell *et al*, 2002; Hanahan & Weinberg, 2000; Swartz *et al*, 2012). As an evolutional process, the clonal expansion of a cell depends on not only the advantage of mutant genotype but also microenvironment (Fig. 1). Many researches demonstrated that microenvironment is closely relevant with tumour progression, as invasion and metastasis. However, researches on microenvironment have not provided effective targeted points for cancer treatment up to now. In this study, we present the cancer cell hypothesis (CCH) based on the progression of tumour in microenvironment and suggest a general therapy by comparing organic and somatic evolutions.

## Organic and somatic evolution

*Cell microenvironment*

The fitness of somatic cells of metazoan, the ability of proliferation and survival, relies on cell microenvironment besides its own adaptability decided by DNA. Cell microenvironment, as a more general concept than tumour microenvironment, consists of the substances and factors around the cell. Because of the complexity and variability of these substances and factors, it is difficult to study the effect of microenvironment on cell evolution. For simplification, we divide all substances and factors of cell microenvironment into two types as nutriment and poison. Nutriments are substances and factors favourable to the survival of cells, as glucose, oxygen and various necessary factors. Poisons are substances and factors harmful to the survival of cells, as toxic metabolites, exogenous toxins and inhibiting factors.

Normal cells completing the normal physiological function of organs have the same DNA as initial germ cell or harbour some additional neutral mutations which do not influence the fitness. A microenvironment can be evaluated by the survival conditions of normal cells. In this study, we indicate the level of microenvironment with Environment Index ($EI$) which ranges at 0 ~ 1. $EI = 1$ means the best microenvironment where normal cells have the highest fitness. $EI = 0$ means the worst microenvironment where normal cells die. $EI$ is decided by the concentrations of nutriments and poisons in microenvironment. A healthy microenvironment ($EI$ close to 1) needs the coordination of all substances and factors, i.e. each nutriment or poison reaches to the optimal concentration. Each of factors is enough to destroy a microenvironment so the polymorphism of microenvironment increases with the decrease of $EI$. Improving a high $EI$ microenvironment is difficult comparing with destroying it.

In the light of the survival status of normal cells, we qualitatively divide the microenvironment into three areas as healthy, diseased and moribund respectively (Fig. 2). In general, the microenvironments decide the fitness of normal cells and the fitness of normal cells decide the health of an individual organism. Therefore, healthy microenvironment means an individual is healthy and an ill individual means the diseased microenvironment. The *EI* of a microenvironment relies on two factors, local factor, i.e. the interaction between cells and systemic factor, i.e. the efficiency of organs and circulatory systems to transport and process nutriments and poisons effectively.

*Organic evolution*

According to Darwinian Theory, organic evolution is obedient to the rule of natural selection in which individuals adapting to the environment are more likely to pass down their genes. Natural selection retains healthy individuals and eliminates diseased individuals until the childbearing age (young individual), but regardless of whether they are healthy in old age (old individual). Therefore, natural selection improves the health degree of young age rather than agedness for individuals of a species with the evolution of DNA of the germ cell.

The fitness of normal cells of an organism is decided by two factors, their own vitality and the level of microenvironment which relies on the function of organs and circulatory systems and the local interaction of cells. Normal cells are derived from the germ cell and all organs and circulatory systems are also developed from the germ cell, so both factors depend on the DNA of germ cell. Therefore, the organic evolution, i.e. the evolution of DNA of the germ cell, brings about improvement of both factors. The causal chain in organic evolution is: the mutations of DNA of the germ cell → the variations of microenvironment and vitality of normal cells → the changes of fitness of normal cells → selection on healthier individuals.

In a long-term organic evolution, normal cells of individuals increases gradually the fitness in healthy microenvironment, but needs not to increase the fitness in unhealthy microenvironment because corresponding unhealthy young individuals are eliminated by natural selection and old individuals are out of the effect of natural selection (Fig. 2). Normal cells in healthiest microenvironment are close to the maximal fitness because organic evolution has worked for a long time since the generation of metazoan hundreds of millions of years ago. Figure 2 shows qualitatively the change in fitness for normal cell of organisms after a long-term organic evolution.

*Somatic evolution*

Compared with organic evolution, i.e. the evolution of germ cell for hundreds of thousands to billions of years, somatic cells are also obedient to Darwinian evolution which lasts only a lifetime of an organism and the mutations within somatic cell cannot be passed down to progenies (Podlaha *et al*, 2012). Normal cells may be proliferative or dormant according to the local microenvironment relying on the location and the developmental stage of an individual. More mutated cells have lower fitness than normal cells (mutated cell 2 in Fig. 3A) based on the fact that most mutations in mitosis are harmful (Eyre-Walker & Keightley, 2007). However, few particular cells, which are sometimes higher in fitness than normal cells, may appear after a large body of mutations. A mutated cell with high fitness may proliferate beyond the procedure of normal development. However, such a cell exceeding normal cells in fitness only appears when microenvironment is unhealthy as diseased (mutated cell 1 in Fig. 3) because normal cells almost have the highest fitness in healthy microenvironments as a result of organic evolution. Thus, there must be a critical point for a mutated cell (point $C$ in Fig. 3A), which divides the microenvironment into two areas. In low $EI$ microenvironment (the left area of point C in Fig. 3A), the mutated cell is higher in fitness than normal cells. In high $EI$ microenvironment (the right area of point C in Fig. 3A), the mutated cell is lower in fitness than normal cells.

*Proliferation rate of mutated cell*

The proliferation rate of normal cells, which complies with the normal development procedure of an organism, can be used as a benchmark for the proliferation of mutated cell. With this benchmark, the proliferation rate of a mutated cell depends entirely on whether it exceeds normal cells in fitness. Therefore, a mutated cell may proliferate faster to form a hyperplasia only when it is locally higher in fitness than normal cells (the colour region in Fig. 3B). The proliferation rate of the mutated cell decreases if the microenvironment is improved to the critical point (region $E_1$ in Fig. 3B) or deteriorates to moribund (region $E_3$ in Fig. 3B). The unhealthy microenvironment (colour region in Fig. 3B) is the risk condition of tumour progression for an organism, through comparing cellular variations in organic evolution and somatic evolution. Many researches on somatic evolution show the evolutionary nature of cancer, in which mutations confer on cancer cells to proliferate out of control (Forbes *et al*, 2008; Podlaha *et al*, 2012). However, the long-term influence of organic evolution on the fitness of in vivo cells is also important for understanding cancer.

## Cancer cell hypothesis

A mutated cell may proliferate to form a tumour in special microenvironment according to organic and somatic evolutions. This mechanism of cell evolution can account for why abnormal and normal cells coexist in an organism. However, cell evolution process does not show why some mutated cells proliferate to benign tumours and others proliferate to malignant tumours featured by invasion and metastasis. What causes the malignance of tumour?

*"Virtuous cell" and "vicious cell"*

Cell evolution depends on microenvironment according to natural selection and cells affect microenvironment in turn. Different cells, which harbour different mutations, may be beneficial or harmful to microenvironment. Thus, we divide cells into two types: "vicious cell" which is harmful to microenvironment and "virtuous cell" which is beneficial to microenvironment. The harm of vicious cells to microenvironment may be derived from some behaviours, such as they produce more metabolic wastes or factors to inhibit other cells. The benefit of virtuous cells for microenvironment may be derived from other behaviours, such as they remove poison or produce more beneficial factors. Obviously, vicious cells are harmful to each other and virtuous cells are favourable to each other.

Virtuous cells are advantageous if closing to each other because they are favourable to each other. A microenvironment, in which the same virtuous cells get together and isolate from other cells, is more beneficial for cell survival. Therefore, autonomous systems, featured as clustered and autocephalous group of the same virtuous cells and isolate from other cells, are advantageous in somatic evolution. Virtuous cells in an autonomous system are high in fitness, so the individual is healthy. Thus, autonomous systems of virtuous cells are also advantageous in organic evolution. All organs of metazoans can be considered as autonomous systems formed in long-term organic evolution.

The microenvironment of a specific location is only friendly to specific cells but hostile to other cells. Cells will not survive in an inappropriate microenvironment due to lack in necessary cell autonomous survival signals (Morrison & Spradling, 2008). The mechanism of anoikis has been considered to prevent normal cells from leaving their original environment and seeding at inappropriate locations (Chiarugi & Giannoni, 2008). Relevant researches suggest also that most kinds of normal cells are virtuous cells, because only virtuous cell is advantageous in both somatic and organic evolutions.

Vicious cells, if they flock together, are disadvantageous because they are harmful to each other. A microenvironment, in which the same vicious cells get together and isolate from other cells, is disadvantageous for cell survival. The nature of vicious cells decides them to tend to expand dispersedly which is contrary to the behaviours of virtuous cells. Therefore, vicious cell is featured as that easy to leave the original location as the result of somatic evolution.

*Cancer cell hypothesis*

Normal cells, consisting of many types of cells, are all derived from the original germ cell. In long-term organic evolution, each type of normal cells, which stay at a special location and complete corresponding physiological function, improve gradually their fitness in special microenvironment and help the individual organism to adapt increasingly to environment. A mutated cell is unlikely higher in fitness than local normal cells if the microenvironment is healthy (Fig. 3A). However, a mutated cell may proliferate faster, i.e. obtains a higher fitness, than normal cells to form a tumour if the microenvironment becomes diseased. A "virtuous" mutated cell with a high fitness as normal cells will develop a benign tumour according to its features. In contrast, a "vicious" mutated cell with a high fitness as normal cells will proliferate to a malignant tumour, featured by invasion and metastasis, because these cells are harmful to each other and other adjacent cells. Therefore, we assume the *cancer cell hypothesis* (CCH) that cancer cell is a mutated cell which is higher in fitness than normal cells in a diseased microenvironment and is harmful to each other and other adjacent cells (destroying microenvironment) by some behaviours, such as producing more poison in metabolism than normal cells.

## Metastatic progression of tumour

*Multistage growth of tumour*

Researches have shown that the vast majority of tumours are monoclonal growths descended from single progenitor cells (Thompson *et al*, 1996; Weinberg, 2007). By theoretical analysis, two conditions determine the progression of a tumour: a mutated cell with higher fitness than normal cells and a given microenvironment where the *EI* is below the critical point for the mutated cell (seed and soil).

In the early phase, when a tumour is small and invisible, the surrounding microenvironment is mainly decided by ambient substances and factors obeying the rule of diffusion. Although they damage microenvironment according to CCH, cancer cells influence the microenvironment little because of too few in number. Therefore, the growth rates for tumours in the early phase are indeterminate because of the diversiform ambient conditions, although the tumour is difficult to be observed due to the small size (Fig. 4).

In the middle phase, a tumour gradually grows up and becomes visible. The effect destroying the microenvironment for cancer cells exceeds the role of ambient conditions when the tumour grows beyond a certain volume. As a result, the diseased microenvironment which is necessary for the tumour to grow (Fig. 3) can self-sustain by enough cancer cells and ambient conditions provides only a secondary influence on microenvironment. Thus, the level of microenvironment depends mainly on the volume of tumour. This correlation between microenvironment and tumour volume answered why the same kind of tumours in the middle phase grow in similar rate (middle phase in Fig. 4). For example, radiographic studies on human cancer growth rates which are evaluated by the tumour volume doubling time (TVDT) show no big difference between primary tumours and matched metastases or between metastases at different sites (Friberg & Mattson, 1997). Therefore, cancer growth rates at the detectable stage are considered as an inherent property reflecting the provenance of the cancers (Klein, 2009; Kusama *et al*, 1972).

Growth rates can plateau as tumours become large (Finlay *et al*, 1988), where proliferation slows. This deceleration depending on that a large tumour results in nutrient restrictions or poison excess, means a microenvironment close to moribund (Fig. 3 area $E_3$), so cancer cells reduce increasingly the fitness. Blood circulation improves the microenvironment once new blood vessels grow inside a tumour. A tumour with new vessels grows again to break through the bottleneck. A large enough tumour, regardless of primary or metastatic, will damage the whole microenvironment of a patient by transferring poison to the whole body or expending excessive nutrient through blood circulation. Figure 4 shows the different growth rates for a tumour in early, middle and late phases.

*Metastasis*

Invasion and metastasis indicator the malignance of a tumour and cause the death of most patients (Hanahan & Weinberg, 2011; Weinberg, 2007). Understanding the mechanism of metastasis is significant for treatment of terminal cancers. Cancer metastasis involves behaviours of malignant cells as separation, movement, stay, survival and proliferation. These behaviours are assumed based on various molecular mechanisms. For example, epithelial-mesenchymal transition is considered as a possible means by which transformed epithelial cells can acquire the abilities to invade and disseminate (Klymkowsky & Savagner, 2009; Polyak & Weinberg, 2009; Thiery *et al*, 2009; Yilmaz & Christofori, 2009). In addition, invasion and metastasis of cancer are relevant with the microenvironment around the tumour involving stromal cells and macrophages (Joyce & Pollard, 2009; Kalluri & Zeisberg, 2006; Kessenbrock *et al*, 2010; Qian & Pollard, 2010). However, the basic mechanism of metastasis is still dubious so far.

According to CCH, cancer cells as "vicious cell" are harmful to not only adjacent normal cells but also other cancer cells. Malignant tumours are loose instead of compact structures because of the nature of "vicious cell". Therefore, cancer cells are lower in adhesion and easier to be separated from a tumour as a result of evolution. In addition, whether cancer cells can move a long distance to the targeted tissue and surviving to proliferate relies on two conditions: the whole body microenvironment increasingly deteriorates and more cancer cells are separated from the primary tumour with the growth of tumour. Obviously, the probability of metastasis increases with the progression of a tumour.

This theoretical result accords with the linear progression model of metastasis in which metastasis is correlative with tumour volume. Contrary to the linear progression model, the parallel progression model of metastasis persists that metastases must be initiated long before the first symptoms appeared or the primary tumour was diagnosed and proliferate in parallel with carcinoma in situ (Collins *et al*, 1956; Friberg & Mattson, 1997). Because of the similar growth rates based on no big different TVDTs, metastases were simply too large to be accounted for by initiation at a late stage of primary tumour development (Klein, 2009). However, the growth rates for tumours in the early phase, which are different from the relatively stable growth rates for tumours in the middle phase, are indeterminate because of the diversiform microenvironment (Fig. 4). Metastases initiated at the late phase of a primary tumour might develop faster because of the diseased microenvironment. Therefore, the parallel progression model is difficult to replace the linear progression model.

## Inspiration for treatment

Many therapeutic methods of cancer have applied in the clinic today. Physical therapies as surgical operation and radiation therapy are suited only to treat preinvasive carcinomas but are helpless for metastatic carcinomas. Chemotherapy and targeted therapy are the treatment approaches for cancers in each stage.

*Chemotherapy*. Chemotherapy as a common therapy for cancers has come into use since the last century (DeVita & Chu, 2008). Various chemotherapy drugs, as antimetabolites, mitotic inhibitors and anti-tumour antibiotics, are widely used to treat cancers. Chemotherapy kills cancer cells by aiming at their general characteristics as rapid division. However, chemotherapy drugs are also toxic to normal cells and harmful to the whole microenvironment of patients (red arrow in Fig. 5). The risk for patients lies in that if chemotherapy cannot kill every cancer cell those residual ones will proliferate and metastasize faster because chemotherapy drugs caused the deterioration of microenvironment. Many cases confirmed that a recrudescent cancer after chemotherapy progresses faster.

*Targeted therapy*. Targeted therapy was started in clinical application a dozen years ago. Targeted drugs, designed by the technology of molecule biology, aim at the specific antigens found on the cell surface (monoclonal antibodies) or the target proteins inside a cell (small molecules). For example, imatinib mesylate, as a tyrosine kinase inhibitor targeting abnormal proteins or enzymes that form inside cancer cells and promote uncontrolled growth, is used to treat gastrointestinal stromal tumour and chronic myeloid leukaemia (Druker *et al*, 2006; Talpaz *et al*, 2002). Targeted therapy restrains the proliferation of cancer cells but has no or only little side effect on normal cells (blue arrow in Fig. 5). It is considered that personalized targeted therapy may be the promising therapy in the future. However, targeted therapy faces at least two challenges in treatment of cancer which is the collection of more than 100 diseases (Weinberg, 2007). First, diverse mutations and polymorphic microenvironments lead to heterogeneous cancer cells (Meacham & Morrison, 2013). Current targeted drugs are effective for only limited types of cancer and targeted drugs for most cancers will be still absent in the near future, because it is difficult and expensive to find effective targeted points of cancers and invent corresponding drugs. Second, capricious cancer cells often generate a resistance to drugs so the curative effect of targeted drugs declines soon for most patients.

   ***Inspiration for new therapy.*** From cell evolution model discussed in this study, cancer cells cannot abnormally proliferate in a microenvironment superior to the critical point (Fig. 3). Diseased microenvironment is the precondition for tumour progression. If the microenvironment is superior to the critical point, a tumour in the early phase may stop to grow so cancer is controlled (green arrow in Fig. 5). Although tumours in the early phase are difficult to be found, improving microenvironment may prevent potential cancers.

For grown tumours in middle and late phases, improving microenvironment can increase the fitness of normal cells to resist the harm of cancer cells. Metastases from a tumour in situ cannot stay or proliferate if the microenvironment of targeted location is superior to the critical point. Therefore, improving microenvironments of the whole body can prevent the invasion and metastasis of tumours in some extent. As if this approach is suited to each phase of cancer, however, improving microenvironment does not necessarily come true for a grown tumour because the deterioration of microenvironment caused by cancer cells is not easy to be halted. If that is the case, the key to cure cancer is how to overcome the harm of cancer cells to improve microenvironment.

Some terminal cancer patients self-cured without any treatment. Although such cases happen with only a very small proportion of patients, they are very common among so many patients around the world. What killed cancer cells which often escape from the severe hunting down of chemotherapy, radiotherapy and other therapies? That is difficult to be explained by present cancer knowledge. The improvement of microenvironment, which may be caused by the changes in diet, sports and emotion, seems a bit reasonable although direct evidences are absent yet.

## Discussions

Cancer is essentially the result of somatic evolution throughout the lifetime of an organism (Breivik, 2005; Greaves & Maley, 2012; Podlaha *et al*, 2012). Proliferating faster than normal cells means cancer cells are higher in fitness. As a result of organic evolution, normal cells are continuously improved their fitness so they achieve almost the highest fitness in healthy microenvironment today. Therefore, cancer cells, as a mutant of normal cell, are higher in fitness than normal cells in only unhealthy microenvironments. Based on cell evolution, we proposed the hypothesis CCH which cancer cells are advantageous in a diseased microenvironment and can spontaneously destroy microenvironment. CCH shows why some tumours are malignant and others are benignant.

Summarizing this study, cancer progression generally needs three conditions: a series of specific mutations result in cancer cells which are higher in fitness than normal cells, microenvironment becomes diseased and "vicious" cancer cells can maintain a diseased microenvironment favourable to self-expansion. Although this statement does not show a specific mechanism in molecular level, which is often required by current cancer biology, it provides a more general and logic conclusion for most cancers. We are convinced of that deterioration of microenvironment plays key roles in initiating a cancer, which is not inferior to the onset of a cancer cell. Maybe, we cannot consider a cancer in the early phase as a cancer because a cancer cell will not develop to a tumour if the microenvironment does not deteriorate. In a way, sickness leads to cancer rather than cancer leads to sickness. "seed" and "soil" are equally important in tumourigenesis.

Current researches of cancer treatment mainly focus on killing or restraining cancer cells. This study may inspire researchers of cancer to find drugs saving normal cells instead of killing or restraining cancer cells. This kind of therapy is different in mechanism from chemotherapy and targeted therapy (Fig. 5). It may be advantageous in three aspects compared with existing therapies. First, it applies to all types of cancer because therapy aims at the general characteristics of all mutated cells. Second, there is no side effect as chemotherapy because therapy is not harmful but favourable to normal cells. Third, cancer cannot produce resistance to drugs because therapy does not increase the selection pressure for cancer cells. Meanwhile, this kind of therapy will face with some challenges. First, improving microenvironment is not easy to come true in the face of the destruction of cancer cells. Second, to develop new therapies based on the theoretical inference of improving microenvironment needs a great deal of deep researches in clinic. Obviously, to evaluate a healthy microenvironment must be simpler than to identify various cancer cells.

In existing therapies, drugs play the role battling with cancer cells. In therapies of improving microenvironment, normal cells next to tumour are the leading roles combating with cancer cells and treatment only provides an ancillary support. That accords with the philosophy of traditional Chinese medicine, in which treatment lies in assisting the self-healing ability of an organism to get rid of diseases. All doctors would advise people to change life style, involving diet, sports and emotion, to prevent cancer. Maybe, changing life style may be a main treatment method for a cancer patient.

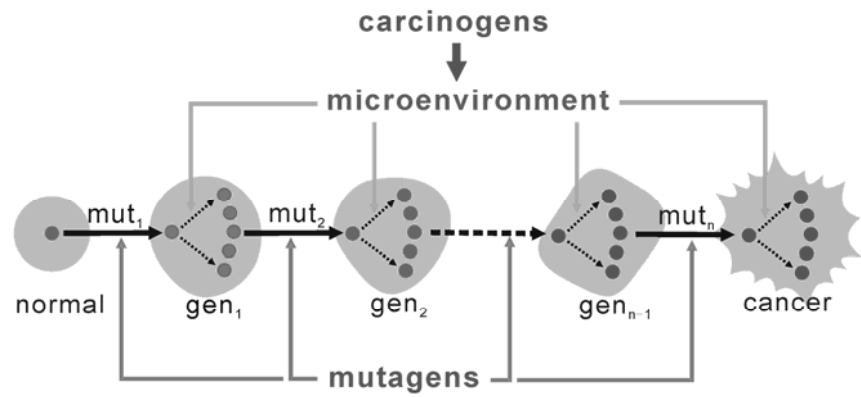

Fig. 1. Multi-step process of tumourgenesis. Each stage in this process involves two events: one cell amid a large cell population sustains a specific mutation and the mutated cell clonally expands. Mutagens are the carcinogenic factors causing the mutation of cell ($mut_1$~$mut_n$) and carcinogens are carcinogenic factors promoting clonal expansion of mutated cell through changing microenvironment.

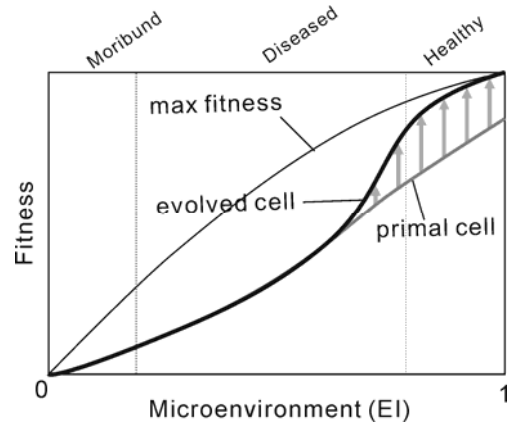

Fig. 2. The fitness of normal cell in organic evolution. The in vivo microenvironment is qualitatively divided into three levels as healthy, diseased and moribund, respectively. The fitness of normal cells decides the fitness of an individual. Long-term organic evolution caused normal cells to have almost the highest fitness in healthy microenvironment. Arrows in the figure indicate the evolution of normal cells which have the same DNA as germ cell.

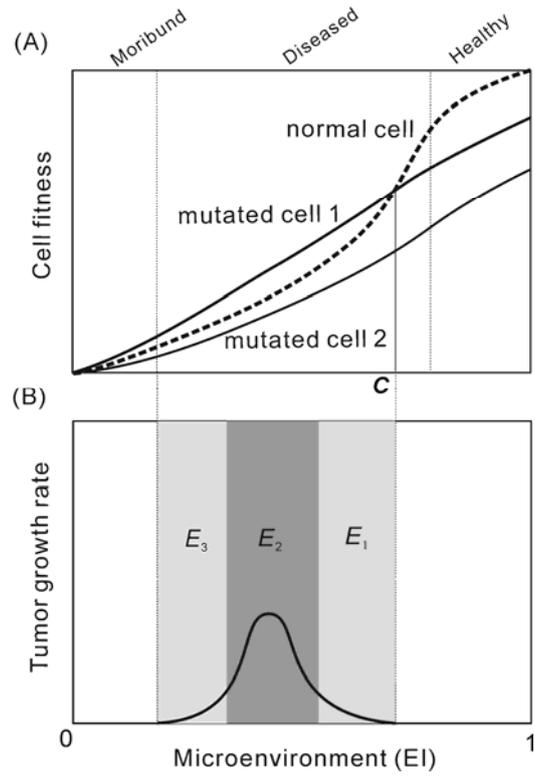

Fig. 3. Mutated somatic cells. Most mutated cells (mutated cell 2) have the fitness below that of normal cells. Few mutated cells (mutated cell 1) have the fitness above normal cells when microenvironment is worse than the critical point (point *C*). Almost all mutated cells have the fitness below normal cells in a healthy microenvironment because normal cells have almost the highest fitness according to organic evolution. Mutated cell 1 may proliferate abnormally, only if it has a high fitness as normal cells. The tumour growth rate increases with the deterioration of microenvironment next to the critical point (area $E_1$). The tumour growth rate decreases with the further deterioration of microenvironment (area $E_3$) because the microenvironment closing to moribund reduces the vitality of all cells. Therefore, a tumour grows with maximum rate in area $E_2$.

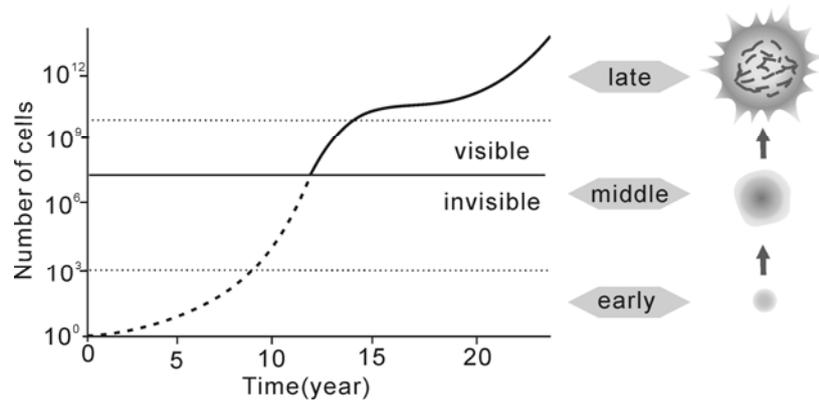

Fig. 4. Three phases of tumour progression. A tumour in the early phase depends entirely on ambient conditions because too few cancer cells cannot influence microenvironment. A tumour in the middle phase grows in a relatively fixed rate because the microenvironment is mainly relevant with tumour volume. A tumour in the late phase grows slow down because a large tumour causes microenvironment close to moribund. Thereafter, newborn blood vessels improve the microenvironment and accelerate tumour growth again.

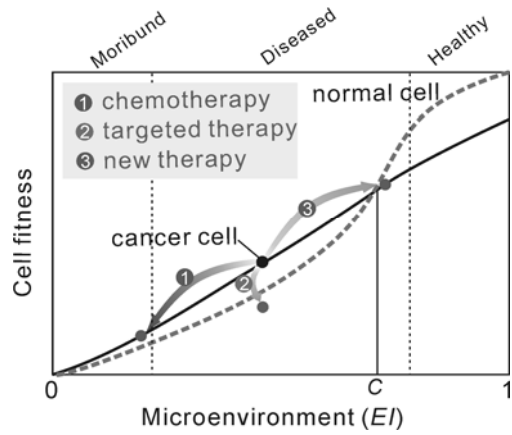

Fig. 5. Therapies. Chemotherapy destroys microenvironment to kill cancer cells so it damages normal cells also. Targeted therapy reduces the fitness of cancer cells by attacking the special targeted point, to halt the progression of a tumour. Targeted therapy has no or only little side effect for normal cells, but cancer cells will generate resistances to targeted drugs. New therapy makes cancer cells lower in fitness than normal cells by improving microenvironment.